\documentclass[nofootinbib,superscriptaddress
]{revtex4}

\usepackage{amsmath}
\usepackage{color}
\usepackage{graphicx}

\def\GeV{{\rm\ GeV}}
\def\MeV{{\rm\ MeV}}
\def\ve{\varepsilon}
\def\la{\lambda}
\def\CG{{\cal G}}
\def\CM{{\cal M}}
\def\CD{{\cal D}}
\def\CA{{\cal A}}
\def\<{\langle}
\def\>{\rangle}
\def\be{\begin{equation}}
\def\ee{\end{equation}}
\def\bea{\begin{eqnarray}}
\def\eea{\end{eqnarray}}

\def\Im{\mathop{\rm Im}\nolimits}
\def\overstar#1{\smash{\overset{*}{#1}}}

\begin{document}

\title{Two photon exchange amplitude with $\pi N$ intermediate states: $P_{33}$ channel}

\author{Dmitry~Borisyuk}
\affiliation{Bogolyubov Institute for Theoretical Physics,
14-B Metrologicheskaya street, Kiev 03680, Ukraine}

\author{Alexander~Kobushkin}
\affiliation{Bogolyubov Institute for Theoretical Physics,
14-B Metrologicheskaya street, Kiev 03680, Ukraine}
\affiliation{National Technical University of Ukraine "KPI",
37 Prospect Peremogy, Kiev 03056, Ukraine}

\begin{abstract}
 We consider two-photon exchange (TPE) in the elastic electron-proton scattering
 and evaluate the effect of $\pi N$ (pion + nucleon) intermediate hadronic states.
 Among different $\pi N$ states, we concentrate on the $P_{33}$ channel;
 thus we effectively include $\Delta(1232)$ resonance with realistic width and shape
 and corresponding background as well.
 In agreement with the previous result, obtained for the zero-width resonance,
 we observe that the TPE correction to the electric form factor is the largest one;
 it grows with $Q^2$ and at $Q^2 \gtrsim 2.5\GeV^2$ exceeds the corresponding elastic contribution.
\end{abstract}

\maketitle

\section{Introduction}
Elastic electron scattering is the
primary tool for measuring the nucleon electromagnetic form factors
which, in turn, reflect the internal structure of the nucleon and
dynamics of the strong interaction inside it.
There are two main methods for measuring the form factors: standard Rosenbluth separation method
and modern polarization transfer method (used since 1998 \cite{Milbrath}).
It was realized about ten years ago that these two methods are in serious disagreement
with respect to the proton form factor ratio $G_E/G_M$ at high $Q^2$ \cite{HowWell}.
While the Rosenbluth method indicated that $G_E/G_M \sim {\rm const}$, 
the polarization data suggested that the ratio decreases almost linearly with $Q^2$.
This disagreement was further confirmed with increased precision experiments,
using both Rosenbluth \cite{LT} and polarization \cite{PT} techniques.
It was suggested that the discrepancy may be due to data analysis done in the Born approximation,
thus leaving out non-trivial higher-order terms, such as two-photon exchange (TPE).
Indeed, first calculations had shown \cite{BMT}, that the discrepancy is, at least partially,
resolved by including TPE corrections.

Besides that, there is now much experimental activity, aimed at the direct observation
of TPE effects in the $ep$ scattering. This includes experiments which are
already completed \cite{Exp1}, in progress \cite{Exp2}, approved or proposed \cite{Exp3}.
A comprehensive review on experimental and theoretical studies of TPE in $ep\to ep$
and other processes, along with bibliography, can be found in Ref.~\cite{Review}.

There are two mainline approaches to the theoretical evaluation of the TPE amplitude:
"quark" and "hadronic" ones.
In the "quark" approach, as its name suggests, the nucleon is viewed as an ensemble of quarks (partons),
interacting according to QCD \cite{GPD,ourQCD,theirQCD,SCET}.
Naturally, the applicability of this approach is limited to the high-$Q^2$ region.
Despite all its advantages, the serious drawback is that it is hard to calculate
the TPE correction to the electric form factor $G_E$ in this approach, while this is surely needed
for the correct interpretation of $G_E/G_M$ measurements.

In the "hadronic" approach TPE is mediated by the production of virtual hadrons
and/or hadronic resonances. The TPE amplitudes are broken into different contributions
according to the intermediate state involved.
The most important and well-established one is the elastic contribution, which corresponds
to pure nucleon intermediate state.
In turn, all other contributions are called inelastic.
Among them, the contributions of some prominent resonances
[$\Delta(1232)$ and others] were studied in Refs.~\cite{BlundenDelta,BlundenRes,ourDelta}. 
In Refs.~\cite{BlundenDelta,BlundenRes} it was shown that their overall effect
{\it on the cross-section} is smaller than that of the elastic contribution,
with $\Delta(1232)$ yielding its main part and the contributions of other resonances partially cancelling each other.

Later, it was found \cite{ourDelta} that $\Delta(1232)$ yields relatively large correction to the $G_E/G_M$ form factor ratio
at high $Q^2$ (far exceeding that of the elastic intermediate state), and that the correction grows with $Q^2$.
This result suggests that the contributions of other inelastic states may also be important
and at least should be estimated carefully.
Unfortunately, all the above-mentioned papers use "zero-width" approximation, i.e. widths of resonances
are assumed to be negligibly small.
This approximation seems rather crude, especially for $\Delta(1232)$, since its width ($\Gamma_\Delta\sim 110\MeV$)
is comparable to the distance from the threshold ($M_\Delta - M - m_\pi \sim 160\MeV$).

To overcome this issue, in the present paper we estimate the inelastic contribution to the TPE amplitude,
arising from the $\pi N$ (pion+nucleon) intermediate states. This may be viewed as a significant improvement
of the previous "resonance" calculations, since most resonances have dominant $\pi N$ content.
Consequently, the advantages of our approach are
\begin{itemize}
\item automatically having correct resonance width
\item automatically having correct resonance shape
\item including not only resonances but background as well
\end{itemize}
The $\pi N$ contribution may further be split into the contributions of different partial waves of the $\pi N$ system.
Though, in principle, all partial waves may be taken into account in our method,
it is particularly useful for the $P_{33}$ channel, where $\Delta$ resides.
The $\Delta$ resonance has almost 100\% $\pi N$ content, thus we will get pure improvement w.r.t. previous works.
The situation is not so simple for other resonances, such as $S_{11}$ and $D_{13}$,
since they have significant $\pi\pi N$ branching ratio;
the corresponding contribution will be missing in the present approach.

Only $P_{33}$ channel will be considered in full detail further.

\section{Background}

The idea of the present calculation is the following. The $\pi N$ system is fully described by its isospin, spin-parity, and invariant mass.
No other internal quantum numbers exist. Thus, with respect to the calculation of the TPE amplitudes,
the $\pi N$ system in the intermediate state is fully equivalent to the single particle with the same isospin, spin-parity and mass
(and properly defined transition amplitudes).
If we are able to calculate the TPE contribution of the resonance with given quantum numbers,
we can do precisely the same thing for the $\pi N$ system of fixed invariant mass and then integrate over invariant masses.

Specifically, if, for the zero-width particle $R$ (resonance) with mass $M_R$
and $R \to \gamma^* N$ transition form factor $A^R(q^2)$ we have
\be \label{genTPE}
 \delta\CG^R = \delta\CG [M_R, A^R(q^2)] \equiv \int A^R(q_1^2) A^R(q_2^2) {\cal K}(M_R,q_1^2,q_2^2) dq_1^2 dq_2^2
\ee
(where $\delta\CG$ stands for any TPE amplitude, and ${\cal K}$ is some kernel, irrelevant for the following discussion),
then the full contribution of the $\pi N$ partial wave with the same quantum numbers will be
\be \label{genTPEpiN}
 \delta\CG^{\pi N} = \int \delta\CG [W, A^{\pi N}(q^2,W)] dW^2,
\ee
where the integration variable $W$ is the invariant mass of the $\pi N$ system and $A^{\pi N}$ is appropriately defined transition form factor.
Note that $q^2$ stands for the square of virtual photon momentum
and should not be confused with the total momentum transfer in the elastic process, $Q^2$.

For the $P_{33}$ channel, all needed formulae [Eq.(\ref{genTPE})] are already derived \cite{ourDelta}.
All we have to do is to establish a correspondence between the transition form factors, used in Ref.~\cite{ourDelta},
and the multipole amplitudes for the production of the $\pi N$ system.

The pion electroproduction is commonly described by the multipole amplitudes $E_{l\pm}$, $M_{l\pm}$, and $S_{l\pm}$,
which are functions of $q^2$ and $W$; the subscript $l$ is pion orbital quantum number and $\pm$ indicates
the total angular momentum $j = l \pm 1/2$.
One can also define helicity amplitudes as follows (see e.g. \cite{AB,MAID}):
\bea \label{mult2AH}
 A_{1/2} &=& -\frac12 \left[ (j+1/2\pm 1) E_{l\pm} \pm (j+1/2\mp 1) M_{l\pm} \right], \nonumber \\
 A_{3/2} &=& \frac12 \sqrt{(j-1/2)(j+3/2)} \left[ \pm E_{l\pm} - M_{l\pm} \right], \\
 S_{1/2} &=& -\frac{1}{\sqrt{2}} (j+1/2) S_{l\pm}.  \nonumber
\eea

Thus we have three "flavours" of the transition amplitude: $A^R$ for the single narrow resonance,
$A$ for the $\pi N$ system, and $A^{\pi N}$ for the "effective resonances" describing $\pi N$ continuum.
As one can guess, the amplitudes $A^{\pi N}$, which are to be put into Eq.~(\ref{genTPEpiN}),
can differ from $A$ [Eq.~(\ref{mult2AH})] only in overall $q^2$-independent factor,
which arises from different normalizations of resonance and $\pi N$ states.
Thus, the factor can easily be determined by considering forward kinematics,
where the imaginary part of the TPE amplitudes is related to the cross-section via the optical theorem.
Nevertheless, a straightforward calculation (in arbitrary kinematics) is  certainly possible,
and to be rigorous, we perform such calculation in the Appendix.

The sought relation is
\be \label{AHfactor}
  A_H^{\pi N} (q^2,W) = A_H(q^2,W) \sqrt{\frac{2W r (2j+1)}{M(W^2-M^2)}},
\ee
where $M$ is the nucleon mass, $r$ is the pion momentum in the $\pi N$ c.m.s.,
and we use the shorthand $A_H$ for any of $A_{1/2}$, $A_{3/2}$, $S_{1/2}$.
Now the argument from the beginning of this section applies,
and the TPE amplitudes will be given by Eq.~(\ref{genTPEpiN}).
The apparent dimension mismatch between $A_H^{\pi N}$ and $A_H$ is not an error, recall that
the dimension of $A_H$ is $\GeV^{-1}$, the dimension of $A_H^R$, meant in Eq.~(\ref{genTPE}), is $\GeV^{-1/2}$,
and Eq.~(\ref{genTPEpiN}) contains additional integration over $dW^2$.

For actual calculation of the TPE amplitudes we wish to employ the {\tt TPEcalc} program \cite{TPEcalc}.
However, both {\tt TPEcalc} and Ref.~\cite{ourDelta} use not $A_H$,
but covariant form factors $F_{1,2,3}$ to describe nucleon-resonance transition.
The latter are related to the transition current matrix element for $R \to \gamma^* N$ as%
\footnote{In arXiv:1209.2746v1, the sign of the $p_\nu F_2$ term is incorrect.}
\be
  \< N|J^\mu|R \> = \frac{1}{4 M^2 \sqrt{MW}}
  ( g^{\mu\alpha} q^\nu - g^{\mu\nu} q^\alpha ) \,
    \bar U \left[ 
        (\hat p \gamma_\nu - p_\nu ) F_1 - p_\nu F_2 + q_\nu F_3
    \right] \gamma_5 V_\alpha,
\ee
where $p$ and $q$ are resonance and photon momenta,
$U$ and $V_\alpha$ are nucleon and resonance spinors,
the states $|N\>$ and $|R\>$ are normalized to unity and the resonance mass is taken to be $W$.

The relationship between $A_H$ and $F_i$ can be obtained using the definitions of $A_H$ (Ref.~\cite{AB}, Eqs.(31-33)) and reads:
\bea \label{AH2Fi}
  K F_1 &=& [(W-M)^2-q^2](A_{3/2} + \sqrt{3}A_{1/2}), \\
  K F_2 &=& [W^2-M^2+q^2](A_{3/2} - \sqrt{3}A_{1/2}) + 2q^2 \frac{W \sqrt{6}}{|\vec q|} S_{1/2}, \nonumber \\
  K F_3 &=&          2W^2(A_{3/2} - \sqrt{3}A_{1/2}) + [W^2-M^2+q^2]\frac{W\sqrt{6}}{|\vec q|} S_{1/2}, \nonumber
\eea
where $|\vec q|$ is photon momentum in the resonance rest frame, and
\be
K = \frac{1}{2M^2}[(W+M)^2-q^2][(W-M)^2-q^2]\sqrt{\pi\alpha\frac{(W-M)^2-q^2}{M(W^2-M^2)}}.
\ee
In full analogy to $A_H$, the form factors $F_i$ should be then "renormalized" according to Eq.~(\ref{AHfactor}),
\be \label{Fifactor}
  F_i^{\pi N} (q^2,W) = F_i(q^2,W) \sqrt{\frac{2W r (2j+1)}{M(W^2-M^2)}}.
\ee
Note that, though for a single narrow resonance $F_i$ are purely real, the calculation of the TPE amplitudes
is well possible when they are complex; the latter is obviously the case here, since the amplitudes $A_H$ are complex.

\section{Technical details}\label{sec:tech}

The multipole amplitudes $E_{l\pm}$, $M_{l\pm}$, and $S_{l\pm}$ were taken from the unitary isobar model MAID2007 \cite{MAID};
the numerical values were downloaded from the dedicated website \cite{MAIDsite}
for $q^2$ from $0$ to $3 \GeV^2$ in steps of $0.05\GeV^2$
and $W$ from $1082$ to $1550 \MeV$ in steps of $15\MeV$.
Note that the site gives the amplitudes for the isospin channels, named $A_p^{(1/2)}$, $A_n^{(1/2)}$ and $A^{(3/2)}$.
The amplitudes for $\gamma^* p \to \pi N (I=3/2)$, which we need here, are $\sqrt{2/3}A^{(3/2)}$.
Then, for each discrete $W$ value $W_i$, the multipole amplitudes were converted into the helicity amplitudes $A_H$
[Eq.(\ref{mult2AH})], and then into the transition form factors $F_{1,2,3}^{\pi N}$ [Eqs.(\ref{AH2Fi}) and (\ref{Fifactor})].
Much like it was described in Refs.~\cite{ourDelta,TPEcalc}, the resulting form factors were fitted with the sum of poles:
\be
 F(q^2,W_i) = \sum_{j=1}^7 \frac{c_{ij} q^2}{q^2-m_{ij}^2},
\ee
where $c_{ij}$ are complex and $m_{ij}$ are real parameters, with the restrictions $m_{i1} = 0$, $m_{i2} = M+W_i$.
The TPE contributions were then calculated: at first for each individual $W$ value,
following procedures described in Ref.~\cite{ourDelta};
and finally they were integrated over $W$ (with the rectangle method), yielding the total $\pi N(P_{33})$ contribution.

To get an impression of the numerical integration errors, we choose several representative kinematical points,
and for that points, tried to
\begin{itemize}
\item vary sampling step in $W$ (between 5, 10 and 15 MeV)
\item use Simpson's rule instead of the rectangle method
\end{itemize}
In all cases the TPE amplitudes changed by no more than 2-3\%, which is small enough
(note that since we are anyway using an expansion in $\alpha$, no more than $\sim 1\%$ accuracy is needed).

Another source of uncertainty is the choice of the upper integration limit in $W$.
In principle, we should integrate up to infinity, but for a numerical calculation we have to choose some finite value,
and 1550 MeV was used in most calculations.
To probe the error resulting from cutting the integral off,
we try to extend the integration limit from 1550 MeV to 1750 MeV or 2 GeV
(note that the MAID multipoles only exist up to $W = 2 \GeV$).
Again, the change in the TPE amplitudes was no more than 3\%, except in high-$Q^2$ region.
If $Q^2$ is high ($\sim 5 \GeV^2$), then increasing upper integration limit changes the TPE amplitudes by about 5-7\%.
Though such precision is still quite acceptable, this implies that the role of the intermediate states
with higher masses increases with $Q^2$.

\section{Results}

As usual, we describe TPE by three invariant amplitudes (generalized form factors)
$\delta\CG_E$, $\delta\CG_M$, and $\delta\CG_3$.
The corrections to the cross-section or polarization observables can be expressed
in terms of these amplitudes; for all relevant formulae see Refs.~\cite{ourDelta,TPEcalc}.

There are three kinematical regions, where it is interesting to look at newly calculated $\pi N(P_{33})$ contribution:
\begin{itemize}
\item the low-$Q^2$ region, which might affect proton radius extraction
\item the vicinity of the resonance, where the zero-width approximation fails
\item the high-$Q^2$ region, where previous works have revealed the substantial growth of the TPE corrections to the $P_t/P_l$ polarization ratio.
\end{itemize}

\begin{figure}
\parbox[t]{0.49\textwidth}{
  \includegraphics[width=0.48\textwidth]{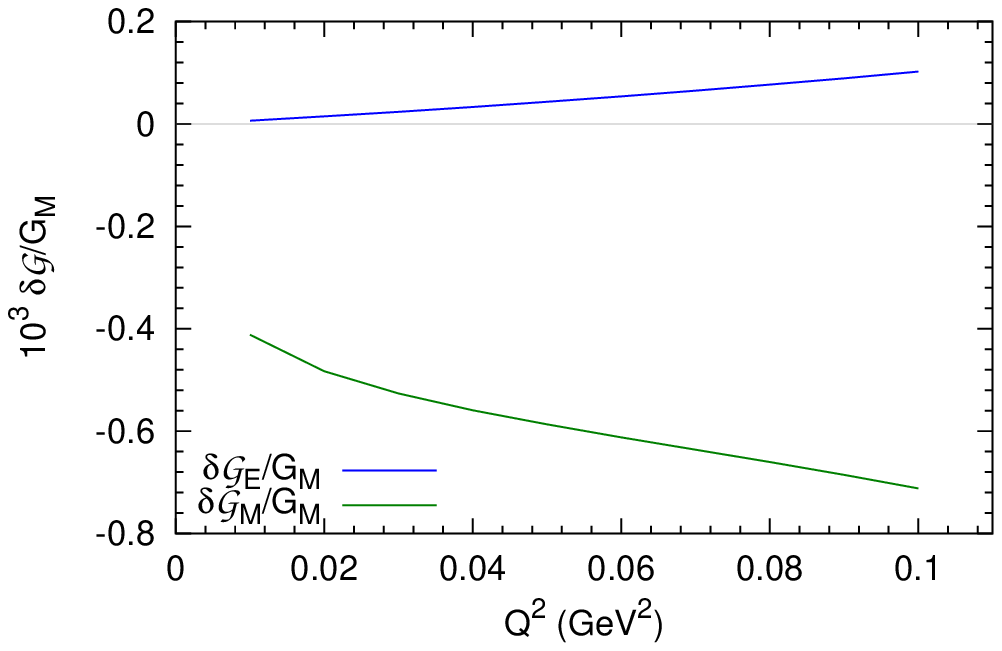}
  \caption{(Color online) The $\pi N$ contribution to the TPE amplitudes at $\ve = 0.25$ and low $Q^2$.}
  \label{fig:lowQ2}
}\hfill
\parbox[t]{0.49\textwidth}{
  \includegraphics[width=0.48\textwidth]{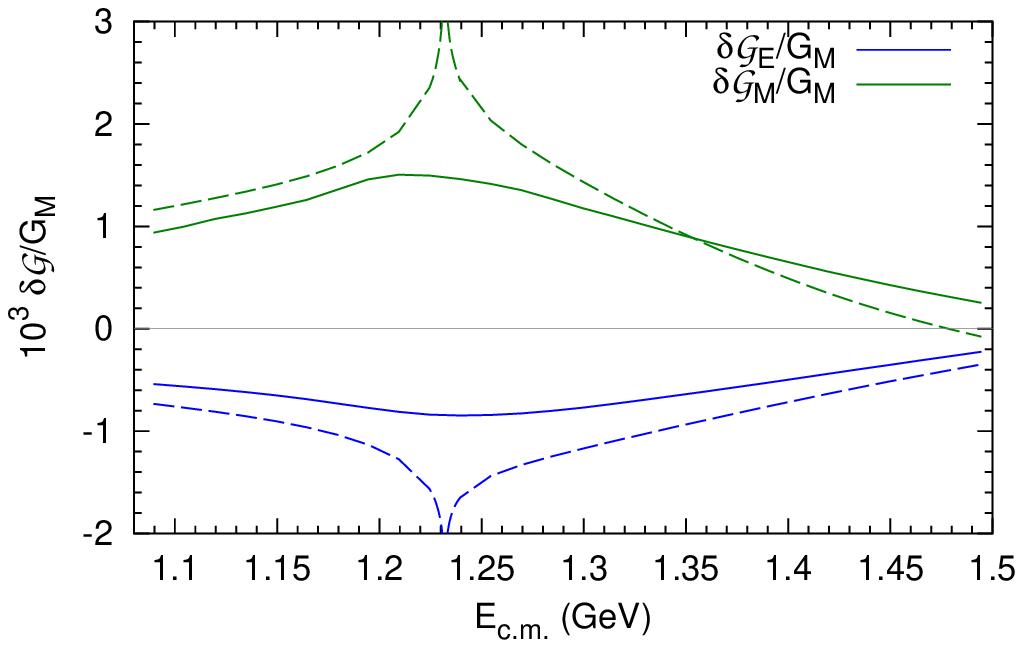}
  \caption{(Color online) The TPE amplitudes near the $\Delta$ resonance, $\theta_{\rm c.m.} = 90^\circ$,
      $\pi N$ contribution from this work (solid), zero-width $\Delta$ \cite{ourDelta} (dashed).}
  \label{fig:res}
}
\end{figure}
At low $Q^2$ the $\pi N$ contributions are small (w.r.t. the elastic contribution)
and change moderately with $Q^2$ (Fig.~\ref{fig:lowQ2}),
thus the proton radius extraction is not affected by the $\pi N$ intermediate states.

The TPE amplitudes in the resonance region are shown in Fig.~\ref{fig:res}.
Just as it was expected, there are smooth bumps at the resonance position, instead of the sharp peaks,
which are seen in the zero-width approximation \cite{ourDelta} (dashed lines).

\begin{figure}[b]
\parbox[t]{0.49\textwidth}{
  \includegraphics[width=0.48\textwidth]{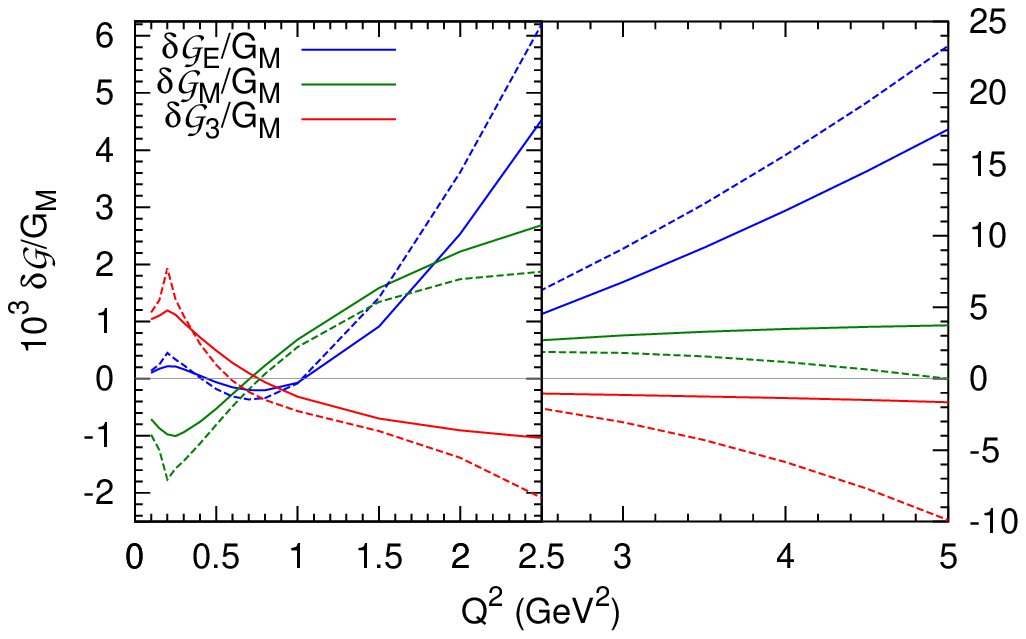}
  \caption{(Color online) $Q^2$ dependence of the $\pi N$ contribution to the TPE amplitudes, at fixed $\ve = 0.25$.
    This work (solid), Ref.~\cite{ourDelta} (dashed).
    Note different $y$-scale in the left and right subplots.}
  \label{fig:Q2}
}\hfill
\parbox[t]{0.49\textwidth}{
  \includegraphics[width=0.48\textwidth]{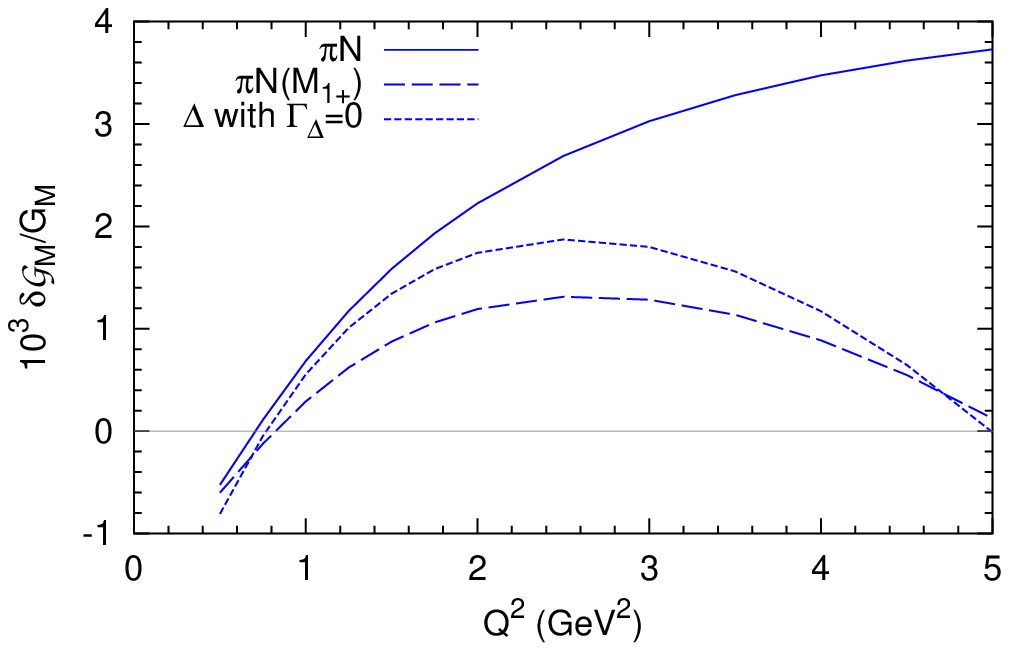}
  \caption{(Color online) The TPE amplitude $\delta\CG_M/G_M$, at fixed $\ve = 0.25$,
    calculated in different approximations.}
  \label{fig:GM}
}
\end{figure}
Figure \ref{fig:Q2} shows $Q^2$ dependence of the calculated TPE amplitudes at fixed $\ve=0.25$ for $Q^2$ up to $5\GeV^2$.
The calculation of Ref.~\cite{ourDelta} is shown with dashed lines for comparison.
We see that for $Q^2$ below $\sim 2.5\GeV^2$ the two approaches give very similar results.
For higher $Q^2$, again in the agreement with Ref.~\cite{ourDelta}, the amplitude $\delta\CG_E$ dominates
and grows with $Q^2$ (though its numerical value is somewhat smaller);
on contrary, the amplitudes $\delta\CG_M$ and $\delta\CG_3$
have quite different values here and in Ref.~\cite{ourDelta}.
When comparing these numerical values, one must keep in mind that present approach differs
from that of Ref.~\cite{ourDelta} in several aspects:
\begin{itemize}
\item now we include $E_{1+}$, $M_{1+}$, and $S_{1+}$ multipole amplitudes,
   whereas Ref.~\cite{ourDelta} effectively includes only $M_{1+}$ amplitude (magnetic transition)
\item the resonance shape differs from pure Breit-Wigner
\item the background contribution is included
\end{itemize}
This is illustrated in Fig.~\ref{fig:GM}, which shows the amplitude $\delta\CG_M$, calculated in three ways:
\begin{itemize}
\item in the present approach,
\item in the present approach with magnetic transition only ($F_2$ and $F_3$ transition form factors set to zero),
\item in the approach of Ref.~\cite{ourDelta}.
\end{itemize}
We see that the difference mainly results from neglecting electric transition in Ref.~\cite{ourDelta}.
In a similar way we have found that the difference of the amplitude $\delta\CG_E$ in the two approaches
is mainly due to neglecting $\Delta$ resonance width.
(Theoretically, the difference also could result from the contribution of states with higher $W$'s,
which are missing in "narrow $\Delta$" calculation.
However, actually this contribution is small, see the end of Sec.~\ref{sec:tech}).

\begin{figure}[t]
\parbox[t]{0.49\textwidth}{  
  \includegraphics[width=0.48\textwidth]{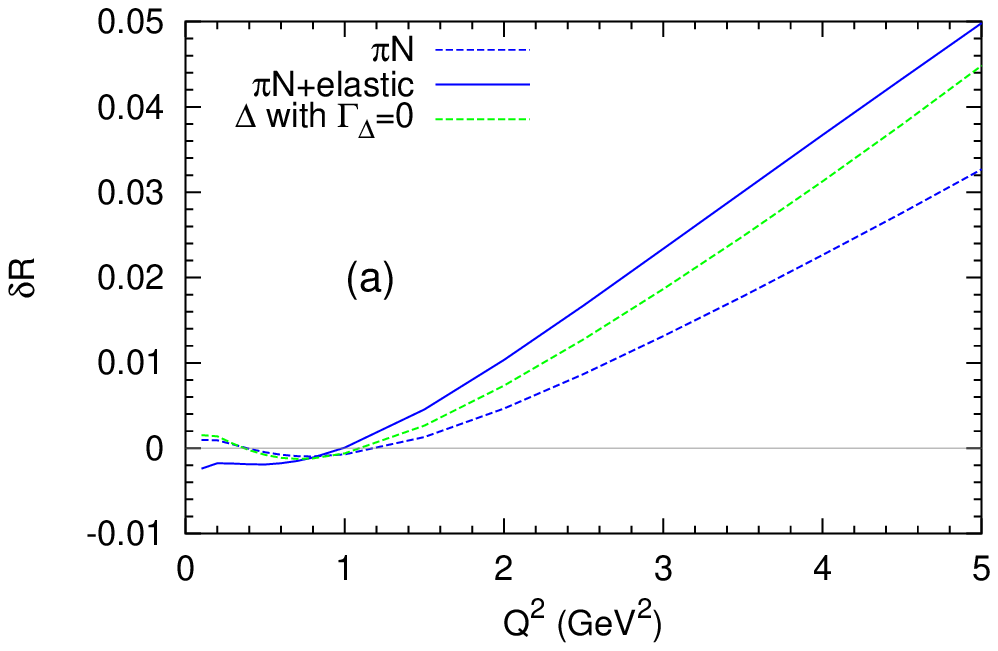}
}\hfill
\parbox[t]{0.49\textwidth}{
  \includegraphics[width=0.48\textwidth]{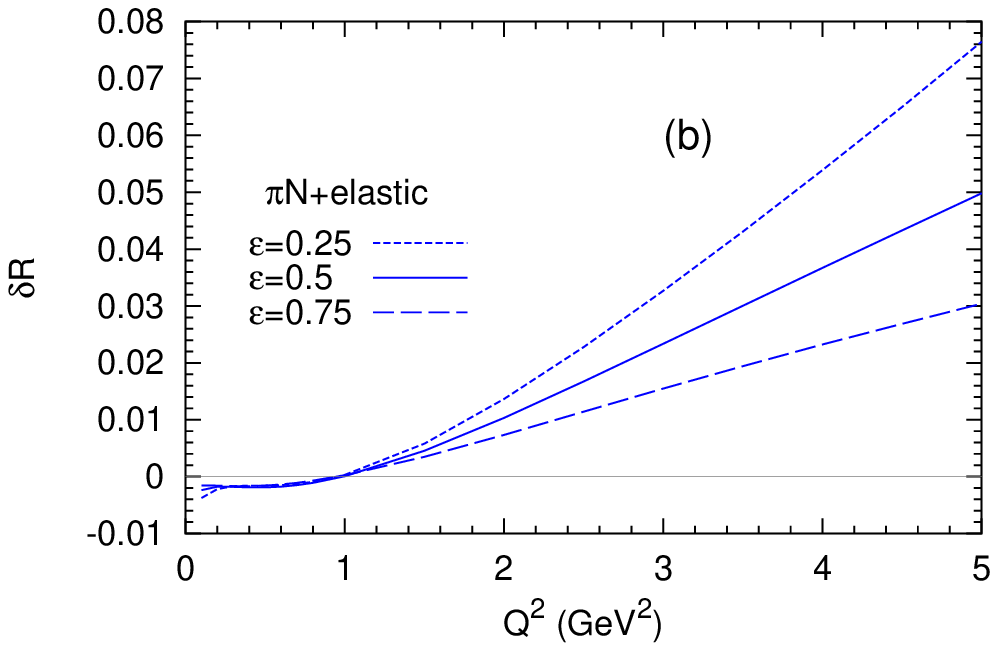}
}
\caption{(Color online) The TPE correction to the proton form factor ratio $R=\mu G_E/G_M$,
  as measured in polarization experiments, various contributions at fixed $\ve = 0.5$ (a)
  and total at different values of $\ve$ (b).
\label{fig:R}
}
\end{figure}
Finally, in Fig.~\ref{fig:R} we plot the TPE correction to the polarization ratio.
At high $Q^2$ we see the same behaviour which was found in Ref.~\cite{ourDelta}, namely
the correction grows rapidly with $Q^2$.
Numerically the correction is $\sim 30\%$ smaller than that obtained in Ref.~\cite{ourDelta}
(for the reasons discussed above).

\section{Conclusions}
We have considered the inelastic TPE contribution, originating from $\pi N$ intermediate states,
and, specifically, the $P_{33}$ partial wave.
Thus we effectively include $\Delta(1232)$ resonance and take into account its finite width.
Numerically we obtain the following results:
\begin{itemize}
\item at small $Q^2$ this contribution is small (negligible w.r.t. the elastic one)
\item the TPE amplitudes have smooth maxima at the resonance position ($E_{\rm c.m.} \approx M_\Delta$)
\item at high $Q^2$ we confirm the findings of Ref.~\cite{ourDelta}, obtained with the zero-width $\Delta$.
  The main correction comes to the generalized electric form factor.
  This correction (and, consequently, the correction to the polarization ratio)
  is relatively large and grows with $Q^2$.
  Its numerical value is somewhat smaller than in Ref.~\cite{ourDelta}
\end{itemize}

In summary, we see, that (contrary to the common belief) the TPE corrections to the polarization ratio
are not negligible at high $Q^2$. 
The question which remains open is: whether the contributions of partial waves other than $P_{33}$
are small or not, how many of them should be taken into account,
and how large is the error, resulting from leaving out partial waves with higher spins.
Surely, it is desirable to answer this question before the TPE corrections are applied to experimental data.
However, this is a separate task, and we plan to do it in further papers.

\begin{acknowledgements}
The authors are grateful to J.~Arrington for useful comments on the manuscript.
\end{acknowledgements}

\appendix
\section{Straightforward calculation of the $\pi N$ contribution via multipoles}

Let the process kinematics be
\be
 e(k) + p(p) \to e(k'') + \pi(r) + p(p'') \to e(k') + p(p'),
\ee
and virtual photon momenta $q_1=k-k''$ and $q_2=k'-k''$.
The imaginary part of the TPE amplitude is given by
\be
 \Im \CM_{fi} = \frac{1}{8\pi^2} \int \frac{(4\pi\alpha)^2}{q_1^2 q_2^2} L_{\alpha\beta} W_{\alpha\beta} d^4\!k'' \delta(k''^2-m^2),
\ee
where $L_{\alpha\beta} = \bar u' \gamma_\alpha (\hat k'' +m) \gamma_\beta u$ is leptonic tensor and
\be
 W_{\alpha\beta} = \sum_h \<p'|J_\alpha|h\>\<h|J_\beta|p\> (2\pi)^3 \delta(p_h-p-q_1)
\ee
is hadronic tensor. Here $J_\alpha$ is electromagnetic current operator,
$|p\>$ and $|p'\>$ are initial and final protons, $|h\>$ is intermediate hadronic state (consisting of one or more particles),
and $p_h$ is its total momentum.
The summation over $h$ actually consists of
\begin{itemize}
\item the summation over different particle types
\item the integration over their momenta (the integration element is $\frac{1}{(2\pi)^3}\frac{d\vec p}{2p_0}$)
\item the summation over their spins
\end{itemize}
This allows us to distinguish between elastic and inelastic contributions,
contributions of various resonances, etc. depending of the nature of the intermediate state $h$.
The contribution of the $\pi N$ intermediate states, which is under consideration now,
has the form
\be
 W_{\alpha\beta} = \int \frac{d\vec p''}{2p''_0}\frac{d\vec r}{2r_0} \frac{1}{(2\pi)^3}
    \sum_{\rm spin} \<p'|J_\alpha|\pi N\>\<\pi N|J_\beta|p\> \delta(p''+r-p-q_1),
\ee
where $p''$ and $r$ are the momenta of the nucleon and pion, respectively.
The tensor $W_{\alpha\beta}$ is convenient to evaluate in the $\pi N$ center-of-mass system, where $\vec p + \vec q_1 = 0$.
Assuming initial and final particles have definite helicities, we may write
\be \label{W_munu}
 W_{\la'\mu';\la\mu} \equiv \overstar\ve_\alpha^{(\la')} \ve_\beta^{(\la)} W_{\alpha\beta} =
   \frac{(8\pi W)^2}{4\pi\alpha} \frac{|\vec r|}{(2\pi)^3 W}
   \int d\Omega \sum_{\mu''} \<\vec q_2\la'\mu'|f^+|\vec r\mu''\>\<\vec r\mu''|f|\vec q_1\la\mu\>,
\ee
where $\ve_\alpha^{(\la)}$ is the polarization vector of virtual photon with helicity $\la$,
$d\Omega$ is pion solid angle, $\mu$, $\mu'$, $\mu''$ are helicities of the initial, final and intermediate protons, and
\be
 \<\vec r\mu''|f|\vec q_1\la\mu\> = \frac{\sqrt{4\pi\alpha}}{8\pi W} \ve_\alpha^{(\la)} \<\pi N|J_\alpha|p\>
\ee
are the helicity amplitudes for the process $\gamma^* p \to \pi N$, defined according to \cite{LL}.
The angular dependence of these amplitudes is determined by the general properties of space rotations;
regardless of the interaction details, they have the following structure \cite{LL}:
\be
 \<\vec r\mu''|f|\vec q_1\la\mu\> = \sum_j \frac{2j+1}{4\pi} \, \CD^j_{-\mu'',\la-\mu}(\vec r) \<\mu''|f^j|\la\mu\>.
\ee
Following \cite{LL}, we use the notation
\be
 \CD^j_{-\mu'',\la-\mu}(\vec r) \equiv \CD^j_{-\mu'',\la-\mu}(\phi,\theta,0)
\ee
for Wigner $\CD$-functions (where $\phi$, $\theta$ are spherical angles of the vector $\vec r$), and
\be
 \<\mu''|f^j|\la\mu\> \equiv \<jm\mu''|f|jm\la\mu\>,
\ee
where $|jm\la\mu\>$ are states with definite angular momentum $j$ and its projection $m$,
and (here and below) $\vec q_1$ is directed along the $z$-axis.
For $\phi=0$, these amplitudes are commonly denoted $H_{1..6}$ \cite{AB}:
\be
 \begin{array}{lclcl@{\qquad}lclcl}
   H_1 &=&  H_{\frac12; -1 \frac12}  &=&  H_{-\frac12;  1 -\frac12}, & H_2 &=& H_{-\frac12; 1 \frac12}  &=& -H_{\frac12;  -1 -\frac12}, \\
   H_3 &=&  H_{\frac12;  1 -\frac12} &=& -H_{-\frac12; -1 \frac12},  & H_4 &=& H_{\frac12;  1 \frac12}  &=&  H_{-\frac12; -1 -\frac12}, \\
   H_5 &=& -H_{\frac12;  0 \frac12}  &=&  H_{-\frac12;  0 -\frac12}, & H_6 &=& H_{\frac12;  0 -\frac12} &=&  H_{-\frac12;  0 \frac12},
 \end{array}
\ee
where
\be
 H_{\mu'';\la\mu} \equiv \<\vec r\mu''|f|\vec q_1\la\mu\>.
\ee
Comparing with Eqs.(7-12) of Ref.~\cite{AB}, we find
\be
 \<\mu''|f^j|\la\mu\> = \frac{4\pi}{\sqrt{2}} \left[ 2\mu \CA^{(2\mu\la)}_{l+} + 2\mu'' \CA^{(2\mu\la)}_{l+1,-} \right],
\ee
where $l=j-1/2$ and
%
\be
 \CA^{(1)}_{l\pm} = -A_{l\pm},
\quad
 (\CA^{(-1)}_{l+}, \CA^{(-1)}_{l+1,-}) = \frac12 \sqrt{l(l+2)} (-B_{l+}, B_{l+1,-} ),
\quad
 (\CA^{(0)}_{l+}, \CA^{(0)}_{l+1,-}) = \frac{\sqrt{-q^2}}{|\vec q|} \frac{l+1}{\sqrt{2}} (S_{l+}, S_{l+1,-})
\ee
(where $A_{l\pm}$, $B_{l\pm}$, and $S_{l\pm}$ are usual multipole amplitudes \cite{AB}).

Now let us proceed with the calculation of $W_{\la'\mu';\la\mu}$.
Switching again to the angular momentum eigenstates according to
\be
 \<\vec q\la\mu|jm\la\mu\> = \sqrt{\frac{2j+1}{4\pi}} \, \CD^j_{\la-\mu,m}(\vec q),
\ee
we have
\bea
 && \int d\Omega \sum_{\mu''} \<\vec q_2\la'\mu'|f^+|\vec r\mu''\>\<\vec r\mu''|f|\vec q_1\la\mu\> = \\
 && = \sum_j \frac{2j+1}{4\pi} \, \CD^j_{\la'-\mu',\la-\mu}(\vec q_2) \sum_{\mu''} \<\la'\mu'|f^{j+}|\mu''\>\<\mu''|f^j|\la\mu\> = \nonumber \\
 && = 4\pi \sum_j (2j+1) \CD^j_{\la'-\mu',\la-\mu}(\vec q_2)
    \left[ (2\mu)(2\mu') \overstar\CA^{(2\mu'\la')}_{l+}\CA^{(2\mu\la)}_{l+} + \overstar\CA^{(2\mu'\la')}_{l+1,-}\CA^{(2\mu\la)}_{l+1,-}  \right] \nonumber 
\eea
and finally
\be \label{W_piN_final}
 W_{\la'\mu';\la\mu} = \frac{(8\pi W)^2}{4\pi\alpha} \frac{|\vec r|}{2\pi W}
   \sum_j \frac{2j+1}{4\pi} \, \CD^j_{\la'-\mu',\la-\mu}(\vec q_2)
   \left[ (2\mu)(2\mu') \overstar\CA^{(2\mu'\la')}_{l+}\CA^{(2\mu\la)}_{l+} + \overstar\CA^{(2\mu'\la')}_{l+1,-}\CA^{(2\mu\la)}_{l+1,-}  \right].
\ee

On the other hand, for the contribution of the infinitely narrow resonance with mass $M_R$,
we will have instead of (\ref{W_munu}), in full analogy with the above,
\bea
 && W^R_{\la'\mu';\la\mu} = \frac{(8\pi W)^2}{4\pi\alpha} \frac{\delta(W-M_R)}{2W}
   \sum_m \<\vec q_2\la'\mu'|f^+|Rjm\>\<Rjm|f|\vec q_1\la\mu\> = \\
 && = \frac{(8\pi W)^2}{4\pi\alpha} \, \delta(W^2-M_R^2) \,
      \frac{2j+1}{4\pi} \, \CD^j_{\la'-\mu',\la-\mu}(\vec q_2) \<\la'\mu'|f^{j+}|R\>\<R|f^j|\la\mu\>, \nonumber
\eea
where photoproduction helicity amplitude is
\be \label{A_R}
 \<Rjm|f|\vec q_1\la\mu\> = \delta_{m,\la-\mu} \sqrt{\frac{2j+1}{4\pi}} \<R|f^j|\la\mu\>
 = \frac{\sqrt{4\pi\alpha}}{8\pi W} \ve^{(\la)}_\alpha \<R|J_\alpha|p\>
\ee
(remember $\vec q_1 \parallel \vec e_z)$ and has the symmetry property
\be
 \<R|f^j|\la\mu\> = \eta_R (-1)^{j-1/2} \<R|f^j|-\!\la-\!\mu\>,
\ee
where $\eta_R$ is resonance parity.
Carefully comparing (\ref{A_R}) with the definition of the standard resonance electroproduction amplitudes
$A^R_{1/2}$, $A^R_{3/2}$ and $S^R_{1/2}$ \cite{AB}, we find
\be
 \<R|f^j|\la\mu\> = \frac{i}{8\pi W} \sqrt{\frac{4\pi}{2j+1}} \sqrt{4M(W^2-M^2)} \binom{2\mu}{1} \CA^{(2\mu\la)}_R,
\ee
where
\be
 \CA^{(1)}_R = A^R_{1/2}, \quad 
 \CA^{(-1)}_R = \mp A^R_{3/2}, \quad
 \CA^{(0)}_R = \mp \frac{\sqrt{-q^2}}{|\vec q|} S^R_{1/2}.
\ee
Upper signs and symbols in the notation like $\binom{2\mu}{1}$ are taken for the resonance parity $\eta_R = (-1)^{j+1/2}$,
and lower ones --- for $\eta_R = (-1)^{j-1/2}$.
Note that overall phase factor is irrelevant. The corresponding hadronic tensor will be
\be
 W^R_{\la'\mu';\la\mu} = \frac{(8\pi W)^2}{4\pi\alpha} \, \delta(W^2-M_R^2) \,
   \frac{2j+1}{4\pi} \, \CD^j_{\la'-\mu',\la-\mu}(\vec q_2)
   \frac{M(W^2-M^2)}{4\pi W^2(2j+1)} \binom{4\mu\mu'}{1}
   \overstar\CA^{(2\mu'\la')}_R\CA^{(2\mu\la)}_R.
\ee
Comparing this with Eq.~(\ref{W_piN_final}), we easily deduce the relation (\ref{AHfactor}).
Indeed, putting in the last equation
\be
 \CA_R = \CA_{j \mp 1/2,\pm} \sqrt{\frac{2W |\vec r| (2j+1)}{M(W^2-M^2)}} ,
\ee
and integrating it over $dM_R^2$, we obtain the term for corresponding spin and parity
from Eq.~(\ref{W_piN_final}).


\begin{thebibliography}{20}
\bibitem{Milbrath} B.D.~Milbrath {\it et al.}, Phys.~Rev.~Lett. {\bf 80}, 452-455 (1998).
\bibitem{HowWell} J.~Arrington, Phys.~Rev.~C {\bf 68}, 034325 (2003).
\bibitem{LT} I.A.~Qattan {\it et al.}, Phys.~Rev.~Lett. {\bf 94}, 142301 (2005).
\bibitem{PT} V.~Punjabi {\it et al.}, Phys.~Rev.~C {\bf 71}, 055202 (2005);
             A.J.R.~Puckett {\it et al.}, Phys.~Rev.~Lett. {\bf 104}, 242301 (2010);
             A.J.R.~Puckett {\it et al.}, Phys.~Rev.~C {\bf 85}, 045203 (2012).
\bibitem{BMT} P.G.~Blunden, W.~Melnitchouk, J.A.~Tjon, Phys.~Rev.~C {\bf 72}, 034612, (2005).
\bibitem{Exp1} M.~Meziane {\it et al.}, Phys.~Rev.~Lett. {\bf 106}, 132501 (2011).
\bibitem{Exp2} A.V.~Gramolin {\it et al.}, Nucl.~Phys.~B (Proc.~Suppl.) {\bf 225-227}, 216 (2012). 
\bibitem{Exp3} M.~Kohl, AIP Conf.~Proc. {\bf 1374}, 527-530 (2011); 
               L.B.~Weinstein, AIP Conf.~Proc. {\bf 1160}, 24-28 (2009); 
               C.F.~Perdrisat, PoS QNP2012 033 (2012).
\bibitem{Review} J.~Arrington, P.G.~Blunden, W.~Melnitchouk, Prog.~Part.~Nucl.~Phys. {\bf 66} 782-833 (2011).
\bibitem{GPD} Y.C.~Chen, A.~Afanasev, S.J.~Brodsky, C.E.~Carlson, M.~Vanderhaeghen, Phys.~Rev.~Lett. {\bf 93}, 122301 (2004).
\bibitem{ourQCD} D.~Borisyuk, A.~Kobushkin, Phys.~Rev.~D {\bf 79}, 034001 (2009).
\bibitem{theirQCD} N.~Kivel , M.~Vanderhaeghen, Phys.~Rev.~Lett. {\bf 103}, 092004 (2009).
\bibitem{SCET} N.~Kivel, M.~Vanderhaeghen, JHEP {\bf 1304}, 029 (2013).
\bibitem{BlundenDelta} S.~Kondratyuk, P.G.~Blunden, W.~Melnitchouk, J.A.~Tjon, Phys.~Rev.~Lett. {\bf 95}, 172503 (2005).
\bibitem{BlundenRes} S.~Kondratyuk and P.G.~Blunden, Phys.~Rev.~C {\bf 75}, 038201 (2007).
\bibitem{ourDelta} D.~Borisyuk, A.~Kobushkin, Phys.~Rev.~C {\bf 86}, 055204 (2012).
\bibitem{AB} I.G.~Aznauryan, V.D.~Burkert, Prog.~Part.~Nucl.~Phys. {\bf 67}, 1-54 (2012).
\bibitem{MAID} D.~Drechsel, S.S.~Kamalov, L.~Tiator, Eur.~Phys.~J.~A {\bf 34}, 69-97 (2007).
\bibitem{TPEcalc} D.~Borisyuk, A.~Kobushkin, arXiv:1209.2746 [hep-ph].
\bibitem{MAIDsite} \url{http://www.kph.uni-mainz.de/MAID/}
\bibitem{LL} V.B.~Berestetskii, E.M.~Lifshits, L.P.~Pitaevskii. Relativistic quantum theory (Oxford, New York: Pergamon Press), 1979.

\end{thebibliography}
\end{document}